\documentclass[pra,nofootinbib]{revtex4}
\usepackage{latexsym}
\usepackage{epsfig}
\usepackage{amssymb}
\newcommand{\beq}{\begin{eqnarray}}
\newcommand{\eeq}{\end{eqnarray}}
\newcommand{\be}{\begin{eqnarray}}
\newcommand{\ee}{\end{eqnarray}}

\begin{document}

\title{Covariant conservation of energy-momentum in modified gravities}
\date{\today}

\author{Tomi Koivisto}
\email{tomikoiv@pcu.helsinki.fi}
\affiliation{Helsinki Institute of Physics,FIN-00014 Helsinki, Finland}
\affiliation{Department of Physics, University of Oslo, N-0316 Oslo, Norway}

\begin{abstract}

An explicit proof of the vanishing of the covariant divergence of the 
energy-momentum tensor in modified theories of gravity is presented. The 
gravitational action is written in arbitrary dimensions and allowed to 
depend nonlinearly on the curvature scalar and its couplings with a scalar 
field. Also the case of a function of the curvature scalar multiplying a 
matter Lagrangian is considered. The proof is given both in the metric 
and in the first-order formalism, i.e. under the Palatini variational 
principle. It is found that the covariant conservation of energy-momentum 
is built-in to the field equations. This crucial result, called the 
generalized Bianchi identity, can also be deduced directly from the 
covariance of the extended gravitational action. Furthermore, we demonstrate 
that in all of these cases, the freely falling world lines are determined 
by the field equations alone and turn out to be the geodesics associated 
with the metric compatible connection. The independent connection in the 
Palatini formulation of these generalized theories does not have a 
similar direct physical interpretation. However, in the conformal 
Einstein frame a certain bi-metricity emerges into the structure of these 
theories. In the light of our interpretation of the independent connection 
as an auxiliary variable we can also reconsider some criticisms 
of the Palatini formulation originally raised by Buchdahl. 

\end{abstract}

\maketitle

\section{Introduction}

Although the Einstein-Hilbert action is the simplest choice producing the 
observational successes of general relativity, no other a priori reason 
prevents from contemplating more general gravitational actions. In fact, in 
some attempts at a more fundamental theory the Einstein-Hilbert action 
can receive corrections characterized by functions of the curvature scalar 
$R$ or other geometrical invariants\cite{Nojiri:2003rz,Mavromatos:2000az}. Such 
modifications of gravity are interesting in cosmology, since they 
could generate the early inflation of the universe
\cite{Mukhanov:1990me}. Moreover, if deviations from the standard general 
relativity become important at low curvature, they could explain the 
current cosmic 
acceleration\cite{Carroll:2003wy, Nojiri:2003ft,Vieira:2003tx,Abdalla:2004sw,
Carroll:2004de,Nojiri:2004fw}. Thus a modification of gravity provides an 
alternative to the cosmological constant or dark energy. 
This is usually considered as a $f(R)$ theory, to which we will mainly 
restrict ourselves here. The Palatini version of these modified gravities 
has also received increasing attention in 
cosmology\cite{Vollick:2003aw,Meng:2003en,Meng:2003ry, 
Kremer:2004bf,Wang:2004pq,Allemandi:2004ca,Allemandi:2005qs,
Koivisto:2005yc,Koivisto:2006ie,Amarzguioui:2005zq,Sotiriou:2005cd}. 

This paper is concerned with the conservation of energy-momentum 
in such modified theories of gravity. In standard general relativity 
energy conservation is built-in to the field equations, since 
the continuity equations (vanishing of the covariant divergence of 
the matter energy-momentum tensor) 
follow directly from the Bianchi identities. When the field equations 
are modified, this is not so easy to see. However, if the modified 
field equations follow from extremization of a covariant action depending 
on the metric and its derivatives, a Noether law for the 
gravitating matter, based on a generalized Bianchi 
identity\cite{Eddington:1987tk,trautmann,Hamity:1992aa,Magnano:1993bd}, 
tells that energy-momentum conservation continues to hold. Nevertheless, 
in section \ref{m_m} we will derive the modified field equations and show 
explicitly that their covariant divergences are equal to zero. This 
provides 
a crucial consistency check for these theories and possibly some 
identities which may be useful in practical calculations. 

In section \ref{m_p} we will apply the Palatini variation to the
corresponding action. This variational principle promotes the connection 
to an independent variable. Interestingly, if and only if the 
Lagrangian is linear in the scalar curvature, the field equations in the 
first order (Palatini) and the standard (metric) formulations coincide. 
The metric variation of a nonlinear gravity theory results in 
fourth order field equations, whereas in the Palatini variation the field
equations turn out to be of second order, since there one can relate the 
curvature scalar $R$ to the trace of the matter energy-momentum tensor to 
eliminate higher derivatives of $R$\cite{Magnano:1995pv}. Thus the 
Palatini variety of these modified gravities is simpler to analyze, and 
may in general exhibit better stability 
properties\cite{Meng:2003ry,Meng:2003uv}. Some possible problems and 
advantages ensuing from assuming the Palatini variational principle 
will be discussed below.  

Although perhaps well-known by specialists in the field, there are aspects 
of geodesics and conservation laws in extended gravities in the first 
order formulation that have not been clearly spelled out in the 
literature, and some confusion has existed. (Note however that these 
results have been presented by Barraco, Guibert, Hamity and 
Vucetich\cite{Barraco:1995aa}, although for only for a specific choice 
of the Lagrangian. Our derivations generalize theirs.)
Furthermore, the essentially problematic role of 
the independent connection in the Palatini  
formulation\cite{Cotsakis:1997cj,Querella:1998ke}, first pointed out in 
the early works by Buchdahl\cite{Buchdahl:1960aa,Buchdahl:1979aa}, has gone 
largely unrecognized in the recent literature\footnote{Recommendable 
discussions are found in Querella's thesis\cite{Querella:1998ke}.}. We 
find it worthwhile to consider in detail some of these rather 
fundamental aspects of the structure of these theories.

The question of whether the covariant energy-momentum conservation holds 
in the Palatini formulation of modified gravity was raised in 
Ref.\cite{Kremer:2004bf}, where it was observed that the continuity 
equation for matter was violated in a perturbative expansion about the 
Einstein-Hilbert action. We are not yet satisfied with the answer given 
in Ref.\cite{Wang:2004pq}. There a suitable auxiliary variable was found 
after successive conformal 
transformations\cite{Allemandi:2004yx,Olmo:2004hj} to show that an 
$f(R)$-theory can be rewritten as a Brans-Dicke theory also in the 
original frame. However, we do not see how this substantiates the 
conservation laws in the Palatini formulation, although the consistency  
of Brans-Dicke theory in the metric formulation is well established (and
follows as a special case of our proof in section \ref{m_m}). In section 
\ref{m_p} we derive the covariant conservation of energy-momentum from the 
field equations in the Palatini formalism. In the following subsection 
we show how it arises from the covariance of the gravitational action. 

In section \ref{i_g} we discuss the physical interpretation and 
consequences of the result. We review the relation between conformally 
equivalent frames in 
extended gravity and consider the geodesic motion of particles. One of 
the main results of this paper is presented here, namely that the 
geodesic hypothesis (that particles follow geodesics of the spacetime 
metric) is unambigiously true in the Palatini form of modified gravity.
We also mention some criticisms raised against this formulation and 
discuss how they limit our possibilities in the interpretation of these 
theories. In section \ref{i_c} we, regarding the attempts to apply 
modifications of gravity as an explanation of the cosmic acceleration, 
point out what seems to be a worthwhile direction to proceed in this 
pursuit in the light of all the results obtained this far.

\section{The conservation law}

\subsection{Metric formalism}
\label{m_m}

We employ natural units, for which $8\pi G=1$. First we will work within 
the metric formalism, in which the standard defitions of the curvature 
variables hold, and the connection is always the Levi-Civita connection of 
the metric $g_{\mu\nu}$. The action we consider is of the form
\be \label{action}
S = \int d^nx \sqrt{-g} \left[\frac{1}{2}f(R,\phi) 
    + \mathcal{L}_\phi(g_{\mu\nu},\phi,\partial\phi) + 
      \mathcal{L}_m(g_{\mu\nu},\Psi,...)\right]. 
\ee     
The two first terms of this action contain nonlinear gravity and 
scalar-tensor theories. The matter Lagrangian $\mathcal{L}_m$ contains 
arbitrary matter fields. The scalar field Lagrangian is written as
\be \label{lag}
\mathcal{L}_\phi = - \frac{1}{2}\omega_{BD} (\phi) (\partial \phi)^2 - 
V(\phi).
\ee 
We have denoted $(\partial \phi)^2 = (\nabla^\alpha \phi) (\nabla_\alpha 
\phi)$. Our results would hold also when more general kinetic terms would 
be included, and in many such cases the scalar field Lagrangian could 
still be 
written in the form of (\ref{lag}) after redefinition of the field. 
Using the definition
\be \label{defi1}
    T_{\mu\nu}^{(i)} \equiv 
-\frac{2}{\sqrt{-g}}\frac{\delta (\sqrt{-g}\mathcal{L}_i)}{\delta(g^{\mu\nu})},
\ee
and calculating the equation of motion for the field $\phi$, one finds that
\be \label{phidiv}
    \nabla^\mu T_{\mu\nu}^{(\phi)} = 
    -\frac{1}{2}\frac{\partial f(R,\phi)}{\partial \phi}\nabla_\nu \phi,  
\ee
regardless of the form of the functions $\omega$ and $V$.

Varying the action (\ref{action}) with respect to the metric we get 
the field equations,
\be \label{fields}
F(R,\phi)R_{\mu\nu}-\frac{1}{2}f(R,\phi)g_{\mu\nu} = 
 (\nabla_\mu \nabla_\nu - g_{\mu\nu} \Box )F(R,\phi) 
 + T_{\mu\nu}^{(\phi)} + T_{\mu\nu}^{(m)},
\ee
where 
\be
F(R,\phi) \equiv \partial f(R,\phi)/\partial R. 
\ee
From now on we lighten the notation by keeping the dependence of $f$ and 
$F$ on $R$ and 
$\phi$ implicit. Taking the covariant divergence on both sides of 
Eq.(\ref{fields}) yields $n$ equations 
\be
(\nabla^\mu F) R_{\mu\nu} + F \nabla^\mu R_{\mu\nu} 
-\frac{1}{2}\left[F \nabla_\nu R 
+ \frac{\partial f}{\partial \phi}\nabla_\nu \phi\right] =
(\Box \nabla_\nu - \nabla_\nu \Box )F + 
\nabla^\mu T_{\mu\nu}^{(\phi)} + \nabla^\mu T_{\mu\nu}^{(m)}.
\ee
These simplify by using Eq.(\ref{phidiv}) and the definition 
$G_{\mu\nu} \equiv R_{\mu\nu}-\frac{1}{2}g_{\mu\nu}R$:
\be
(\nabla^\mu F) R_{\mu\nu} + F \nabla^\mu G_{\mu\nu} 
= (\Box \nabla_\nu - \nabla_\nu \Box )F + \nabla^\mu T_{\mu\nu}^{(m)}.
\ee
On purely geometrical grounds, $\nabla^\mu G_{\mu\nu} = 0$ 
and $(\Box \nabla_\nu - \nabla_\nu \Box )F = R_{\mu\nu}\nabla^\mu F$. 
These identities follow from the definitions of the tensors $G_{\mu\nu}$ 
and $R_{\mu\nu}$\cite{misner}. Therefore $\nabla^\mu T_{\mu\nu}^{(m)}= 0$, 
and the conservation energy-momentum in modified $f(R,\phi)$-gravities in 
the 
metric formulation is confirmed. 

Recently there has been interest in models where a function of the curvature 
scalar enters into the action to multiply a matter 
Lagrangian\cite{Mukohyama:2003nw,Dolgov:2003fw,Nojiri:2004bi,
Inagaki:2005qp,Nojiri:2004fw,Allemandi:2005qs}. 
Such terms were not included in our action (\ref{action}), but we consider
them separately here. We set $f=R$ and $\omega(\phi)=V(\phi)=0$ for 
simplicity, since the forms of these functions are irrelevant here. Thus 
we write 
the action as
\be \label{action2}
S = \int d^nx \sqrt{-g} 
    \left[\frac{1}{2}R + k(R)\mathcal{L}_m(g_{\mu\nu},\Psi,...)\right].
\ee     
The field equations are then 
\be
G_{\mu\nu} = -2K\mathcal{L}_m R_{\mu\nu} 
  + (\nabla_\mu \nabla_\nu-g_{\mu\nu}\Box)2\mathcal{L}_m K + kT^{(m)}_{\mu\nu}, 
\ee
where $K \equiv dk/dR$ and $R$-dependence is again kept implicit. 
Sticking still to the definition in (\ref{defi1}) and proceeding as 
previously, one finds that now
\be \label{mdiv}
k \nabla^\mu T^{(m)}_{\mu\nu} = (\nabla^\mu R)\left[g_{\mu\nu}\mathcal{L}_m 
         - T^{(m)}_{\mu\nu}\right]K.
\ee
If $k$ is a constant or the matter Lagrangian does not explicitly depend on 
the metric, the covariant divergence of the energy-momentum tensor 
vanishes. Otherwise the matter fields must satisfy equations of motions which
are equivalent to (\ref{mdiv}).   

\subsection{Palatini formalism}
\label{m_p}

Next we check the conservation laws under the Palatini variational principle 
which promotes the connection to an independent variable. 
The Ricci tensor is then defined solely by the connection 
$\hat{\Gamma}^\alpha_{\beta\gamma}$,
\be
\hat{R}_{\mu\nu} = \hat{\Gamma}^\alpha_{\mu\nu , \alpha}
       - \hat{\Gamma}^\alpha_{\mu\alpha , \nu}
       + \hat{\Gamma}^\alpha_{\alpha\lambda}\hat{\Gamma}^\lambda_{\mu\nu}
       - \hat{\Gamma}^\alpha_{\mu\lambda}\hat{\Gamma}^\lambda_{\alpha\nu},
\ee 
and $f$ in the action is regarded as a function the metric, the 
connection, and the scalar field,
\be \label{action_p}
S = \int d^nx \sqrt{-g} 
    \left[\frac{1}{2}f(R(g_{\mu\nu},\hat{\Gamma}^\alpha_{\beta\gamma}),\phi) 
    + \mathcal{L}_\phi(g_{\mu\nu},\phi,\partial\phi) 
    + k(R(g_{\mu\nu},\hat{\Gamma}^\alpha_{\beta\gamma}))
      \mathcal{L}_m(g_{\mu\nu},\Psi,...)\right]. 
\ee     
In the first order formalism, the field equations got by setting variation 
with respect to the metric to zero seem simple,
\be \label{fields2}
F \hat{R}_{\mu\nu}-\frac{1}{2}fg_{\mu\nu} + 2K\mathcal{L}_m 
\hat{R}_{\mu\nu} = T_{\mu\nu}^{(\phi)} + kT_{\mu\nu}^{(m)}.
\ee
However, now $R \equiv g^{\mu\nu}\hat{R}_{\mu\nu}$ and $\hat{R}_{\mu\nu}$ 
are not the ones constructed from the metric. 
By varying the action (\ref{action_p}) with respect to  
$\hat{\Gamma}^\alpha_{\beta\gamma}$, one gets the condition
\be
\hat{\nabla}_\alpha\left[\sqrt{-g}g^{\beta\gamma}(F+2K\mathcal{L}_m)\right]=0,
\ee
where $\hat{\nabla}$ is the covariant derivative with respect to 
$\hat{\Gamma}$, implying that the connection is compatible with the 
conformal metric 
\be \label{conformal}
h_{\mu\nu} \equiv (F+2K\mathcal{L}_m)^{2/(n-2)}g_{\mu\nu} \equiv 
\omega^{2/(n-2)}g_{\mu\nu}.  
\ee
However, the connection $\hat{\Gamma}^\alpha_{\beta\gamma}$ is not the 
physically interesting connection on the manifold, just as the 
metric $h_{\mu\nu}$ does not have any direct physical content. It just governs 
how the tensor we call $\hat{R}_{\mu\nu}$ appearing in the action settles 
itself in order to minimize 
the action. One could also consider the case that the 
metric $h_{\mu\nu}$ is the measurable, but that
would lead to freely falling particles following geodesics that are not 
those 
corresponding to the metric. This would lead to a different 
theory, which could also been considered\cite{Poplawski:2005sc}), but
will not concern us for now\footnote{A somewhat related issue is that if the 
matter Lagrangian depends on the connection, it must be specified whether 
this connection is the Christoffel one or the one which the Palatini 
variation yields. The 
latter would lead to a complicated theory of the Dirac 
field\cite{Vollick:2004ws}, but could resolve the electron-electron 
scattering problem perhaps present in the former\cite{Flanagan:2003rb}.}.  
We will return to these discussions in more detail below.

As the Ricci tensor is constructed from 
the metric $h_{\mu\nu}$, the easiest way to find it in terms of 
$g_{\mu\nu}$ is to use a conformal transformation. We get
\be \label{riccit}
\hat{R}_{\mu\nu} = R_{\mu\nu} +\frac{(n-1)}{(n-2)}
             \frac{1}{\omega^2}(\nabla_\mu \omega)(\nabla_\nu \omega) - 
             \frac{1}{\omega}(\nabla_\mu \nabla_\nu \omega)
            -\frac{1}{(n-2)}\frac{1}{\omega}g_{\mu\nu}\Box \omega.
\ee
Note that the covariant derivatives above are with respect to 
$g_{\mu\nu}$.
The curvature scalar and Einstein tensor 
follow straightforwardly,
\be \label{riccis}
R = R(g) - \frac{2(n-1)}{(n-2)}\frac{1}{\omega} \Box \omega 
    + \frac{(n-1)}{(n-2)}\frac{1}{\omega^2}(\partial \omega)^2,
\ee
where $R(g)$ is the corresponding scalar constructed from the metric 
$g_{\mu\nu}$,
\be \label{hatg}
\hat{G}_{\mu\nu} =  G_{\mu\nu} + 
             \frac{(n-1)}{(n-2)}\frac{1}{\omega^2}(\nabla_\mu 
               \omega)(\nabla_\nu \omega) 
             - \frac{1}{\omega}\left(\nabla_\mu \nabla_\nu \omega - 
               g_{\mu\nu}\Box\right)\omega
             -\frac{(n-1)}{2(n-2)}\frac{1}{\omega^2}g_{\mu\nu} (\partial 
              \omega) ^2,
\ee 
and a somewhat more tedious calculation\footnote{One can
arrive at this result by relating the $g$-divergence of 
$\hat{G}_{\mu\nu}$ to its vanishing $h$-divergence via the difference of 
the corresponding connection coefficients, or alternatively by taking 
directly the $g$-divergence of Eq.(\ref{hatg}).} then reveals that
\be \label{gdiv}
\nabla^\mu \hat{G}_{\mu\nu} = -\frac{(\nabla^\mu 
\omega)}{\omega}\hat{R}_{\mu\nu}.
\ee  
Taking now the divergence of the field equations (\ref{fields2}) 
similarly as in the previous case, we get 
\be
(\nabla^\mu \omega) \hat{R}_{\mu\nu} + \omega \nabla^\mu \hat{R}_{\mu\nu}
-\frac{1}{2}\left[(\omega-2K\mathcal{L}_m) \nabla_\nu R + 
\frac{\partial f}{\partial \phi}\nabla_\nu \phi\right] =
\nabla^\mu T_{\mu\nu}^{(\phi)} + k\nabla^\mu T_{\mu\nu}^{(m)}
+ (\nabla^\mu k) T_{\mu\nu}^{(m)}.
\ee
This simplifies, by using Eqs.(\ref{phidiv}) and (\ref{gdiv}), to 
\be \label{mdiv2}
k \nabla^\mu T^{(m)}_{\mu\nu} = (\nabla^\mu R)\left[g_{\mu\nu}\mathcal{L}_m 
         - T^{(m)}_{\mu\nu}\right]K.
\ee
These conditions are formally the same as the ones found in the metric 
formulation, Eq.(\ref{mdiv}), but here $R$ is given by Eq.(\ref{riccis}).
Thus the divergence of the matter energy-momentum tensor again vanishes 
identically when $k$ is a constant. This refutes the contrary conclusion
in Ref.\cite{Kremer:2004bf}. If matter is nonminimally coupled to
curvature (i.e. $K \neq 0$), we have $n$ constraints which the matter 
fields must satisfy. These are satisfied identically in the special case 
that $\partial \mathcal{L}_m/\partial g^{\mu\nu} = 0$. Otherwise the 
non-minimal curvature coupling influences the matter continuity 
non-trivially.

\subsection{Generalized Bianchi identity}
\label{m_b}

These results may be understood in light of the generalized Bianchi 
identities\cite{Eddington:1987tk} as a consequence of the Noether theorem.
Trautman has discussed conservation laws in gravitation  
in Ref.\cite{trautmann}. Hamity and Barraco\cite{Hamity:1992aa}
have brought up generalized Bianchi identities in the Palatini 
formulation. Here we follow the Magnano and Sokolowski's 
derivation\cite{Magnano:1993bd}, which is straightforward to generalize 
by considering $n$ dimensions, including scalar field couplings in the 
gravitational action and applying the Palatini variational principle, but 
we outline the procedure here for completeness. Especially the 
incorporation of the independent connection $\hat{\Gamma}$ in this 
derivation might not be immediately clear\cite{Wang:2004pq}. 

Consider an infinitesimal point transformation 
\be \label{ptransf}
x^\mu \rightarrow x'^\mu = x^\mu + \epsilon \xi^\mu,
\ee 
where $\xi^\mu$ is a vector field vanishing on the boundary 
$\partial \Omega$ of a region $\Omega$. The fields entering into the 
gravitational action are shifted such that $f(x) \rightarrow f(x')$. Since 
the gravitational action is extremized in the classical solution, we 
demand that the action (\ref{action_p}) is invariant to first 
order in the infinitesimal parameter $\epsilon$ under the transformation 
(\ref{ptransf}):
\be \label{vari}
0 = \delta S = \int_\Omega d^n x \left( 
\frac{\delta[\sqrt{-g}(\frac{1}{2}f+\mathcal{L}_\phi)]}{\delta g^{\mu\nu}} 
\delta g^{\mu\nu} 
+
\frac{\delta[\sqrt{-g}(\frac{1}{2}f+\mathcal{L}_\phi)]}{\delta 
\hat{\Gamma}^\alpha_{\beta\gamma}} \delta \hat{\Gamma}^\alpha_{\beta\gamma} 
+
\frac{\delta[\sqrt{-g}(\frac{1}{2}f+\mathcal{L}_\phi)]}{\delta \phi} 
\delta \phi\right).
\ee
In fact, $f$ can depend on the metric and its derivatives up to any 
order $m$. Note that for the metric formalism action (\ref{action}) $m=2$, 
whereas for the Palatini formalism action (\ref{action_p}) $m=1$. 
Then one applies the Gauss theorem $m$ times and drops the 
boundary terms to arrive at
\be \label{defi2}
2\sqrt{-g} Q_{\mu\nu} \equiv \frac{\delta(\sqrt{-g}f)}{\delta g^{\mu\nu}} 
\delta g^{\mu\nu}  
\equiv
\frac{\partial(\sqrt{-g}f)}{\partial g^{\mu\nu}}
- \partial_\alpha \frac{\partial(\sqrt{-g}f)}{\partial g^{\mu\nu}_{,\alpha}} 
+ \dots
+ (-1)^m \partial_{\alpha_1} \dots \partial_{\alpha_m}  
\frac{\partial(\sqrt{-g}f)}{\partial g^{\mu\nu}_{,\alpha_1\dots\alpha_m}}.
\ee
Here the important caveat is that we assume all the derivative terms
up to $m$'th order to vanish in the boundary $\partial M$. This condition
is rather problematical, but now we just take for granted that it holds. 
In fact this was done already in the subsection
\ref{m_m} when we wrote the resulting order field equations in 
the metric formalism. This so called Cauchy problem is overcome in the 
Palatini formalism discussed in subsection \ref{m_p}, and in general for 
an action involving just the variables $g_{\mu\nu}$, $\hat{\Gamma}$, and 
possibly their first derivatives. In the metric formalism the second term 
in Eq.(\ref{vari}) identically zero since there is no  
independent connection variable, and in the Palatini formulation the 
extremization of the action with respect to the variations in the connection
$\hat{\Gamma}$ guarantees the vanishing of the second term. Similarly, the 
equation of motion for the scalar field states nothing but that the third 
term 
disappears. Since under the transformation (\ref{ptransf}) 
the metric transforms as $g^{\mu\nu} \rightarrow 
g^{\mu\nu} + \epsilon\nabla^{(\mu}\xi^{\nu)}$, we then get, using the 
symmetry $\mu \leftrightarrow \nu$ and the definitions of 
Eqs.(\ref{defi1}), (\ref{defi2}) that
\be \label{result}
\delta S = 2\epsilon\int_\Omega d^n x \sqrt{-g}(Q_{\mu\nu} - T^{(\phi)}_{\mu\nu})\nabla^\mu\xi^\nu
         = 2\epsilon\int_{\partial\Omega} \sqrt{-g}(Q_{\mu\nu} - T^{(\phi)}_{\mu\nu})\xi^\nu dS^\mu
         - 2\epsilon\int_{\Omega} d^n x \sqrt{-g}\nabla^\mu(Q_{\mu\nu} - T^{(\phi)}_{\mu\nu})\xi^\nu = 0.
\ee
The surface term is zero since $\xi^\nu$ vanishes at the boundary. Since it 
is otherwise arbitrary, $\nabla^\mu Q_{\mu\nu} 
=  \nabla^\mu T_{\mu\nu}^{(\phi)}$. The energy conservation follows
from the field equations since they read 
$Q_{\mu\nu} = T_{\mu\nu}^{(\phi)} + T_{\mu\nu}^{(m)}$.   
Thus the matter action can be considered to be separately invariant 
under the point transformations. 

\section{Discussion}

\subsection{Consequences and interpretations of the result}
\label{i_g}

We have derived the generalized Bianchi identity, which 
states that $\nabla^\mu Q_{\mu\nu} = 0$, where $Q_{\mu\nu}$ is the 
generalized Einstein tensor as defined in Eq.(\ref{defi2}), and explicit 
form of which can be read from Eq.(\ref{fields}) or Eq.(\ref{fields2}). 
From this follows, in the presence of matter fields, that one can 
consistently rely on the vanishing of the covariant divergence of 
the matter energy-momentum tensor. However, if the action integral
(\ref{action}) or (\ref{action_p}) 
includes a product of a function of the curvature scalar and a 
matter Lagrangian, the energy-momentum tensor of the latter is not 
independently conserved except in special cases, but 
instead obeys Eq.(\ref{mdiv}) or Eq.(\ref{mdiv2}). This is not 
a sign of inconsistency, since for example for scalar matter with 
Lagrangian of the form (\ref{lag}), one can readily verify that 
the equations of motion are equivalent to the conditions (\ref{mdiv}) 
(or (\ref{mdiv2})). Coupling a dark component of the 
universe to the curvature in such a way provides an approach to the 
cosmological constant problem\cite{Mukohyama:2003nw,Dolgov:2003fw,
Nojiri:2004bi,Inagaki:2005qp,Nojiri:2004fw, Allemandi:2005qs}. 

Such a coupling appears when one rewrites an extended gravity theory in 
the conformally equivalent Einstein frame,
using the metric $\hat{g}_{\mu\nu} \equiv F g_{\mu\nu}$. As is well known, 
the energy-momentum 
tensor is not covariantly conserved in the conformal Einstein frame.
It is indeed sometimes considered that matter is minimally coupled to 
gravity in the physical frame\cite{Magnano:1993bd}, and that one can and 
should take advantage of this fact when determining which of the 
conformally equivalent metrics is the physical one. While this is not an 
unreasonable assumption, it is neither a compelling argument. In an 
extended theory of gravity of the form (\ref{action}), one generically 
finds a dynamical effective gravitational constant. By transforming to the 
Einstein conformal frame, a constant force of gravity is recovered, but 
the masses of particles are found to evolve in time. Both of these cases 
are mathematically self-consistent, and thus there is no a priori reason 
to exclude the former possibility. One could say that the two frames 
represent different theories, if a theory is understood as a 
specification of the physical variables and an action written in terms 
of them. In different a terminology a theory just equals an action, 
but then (most of) the physics is still left unpredicted by a 
theory.

In the Palatini formulation the conformal metric (\ref{conformal}) inevitably
appears also when considering an extended gravity the Jordan frame. 
According to the least action principle, the gravitational section in the 
action (\ref{action_p}) settles in such a way that a function of the 
contraction of the Ricci tensor associated with the conformal metric 
$h_{\mu\nu}$ is extremized. This Ricci tensor has the usual relation to 
parallel transport in the manifold, but now to the parallel transport 
according to the connection of the Einstein conformal metric. Therefore 
the resulting field equations are different from both standard general 
relativity and its extensions when considered in the metric formalism. 
In the Palatini formulation the response of gravity to matter  
features, effectively, additional sources in the matter sector.
Still, the equations of motion for matter take the same form in all of these 
cases. The energy-momentum conservation laws are the same. The conformal 
metric $h_{\mu\nu}$ plays an interesting mathematical role in the 
Palatini formulation of the action principle, 
suggesting a possible derivation from and an interpretation within a more 
fundamental framework of quantum gravity. However, from the viewpoint of 
the resulting classical theory of gravitation we are now considering, the 
metric $h_{\mu\nu}$ and the connection $\hat{\Gamma}$ associated with it 
can be regarded as just auxiliary fields which were used in the 
formulation of the action principle to derive the field equations.

In fact, conceptual difficulties can arise if these
auxiliary fields are tried to be loaded with more meaning. If one
wants to classify the spacetime according to the connection 
$\hat{\Gamma}$, one then has to ascribe to the variational principle 
the task of selecting a particular spacetime from a variety consisting 
of whole classes of spacetimes, including Riemannian, Weyl and more general 
manifolds. This is true even when considering the Einstein-Hilbert action 
in the Palatini context. The resulting theory is then general relativity 
with its Riemannian geometry, but actually one is implicitly 
dealing with much broader geometrical setting. To some authors, this
has implied a strong objection against the ''Palatini's device'' in 
general\footnote{Objections have been raised against the Palatini approach also
on the grounds of indeterminacies found in specific cases of quadratic 
Lagrangians (when $n=4$). For example, when $f(R) \sim R^2$, the field 
equations ensuing from the Palatini variation determine the metric only up to a 
conformal factor. However, there is no reason to expect that every exotic 
Lagrangian should yield a viable or even well-defined physical theory in 
the Palatini formalism. In the metric formalism one can as well find 
quadratic invariants that lead to field equations which leave the metric 
quite undetermined\cite{Buchdahl:1979aa}. On the other hand, in the 
Palatini formulation of modified gravity the trace equation can admit 
several solutions, but it is not clear if this interesting feature 
should be regarded as a serious drawback of the Palatini approach. For 
example, if $f(R)$ is a polynomial function of the order $m$, there in 
principle can exist $m$ possible roots for the solution.}
\cite{Buchdahl:1960aa,Buchdahl:1979aa,Cotsakis:1997cj, Querella:1998ke}.
However, this issue is simply avoided if one regards the spacetime as  
Riemannian from the beginning. If it indeed is necessary to specify which 
is ''the connection'' on the manifold, it is natural to choose 
also now the Christoffel symbol $\Gamma$ of the metric $g_{\mu\nu}$. There 
is no logical inconsistency in making this prescription a priori and 
without referring to the other independent connection $\hat{\Gamma}$. Just as 
in the metric formulation, the geometrical class of the spacetime is then fixed 
ab initio and not known only after the equations of motion have been solved. 
Nevertheless, the conceptual problem originally pointed out by Buchdahl 
has motivated studies of a so called constrained first order formalism, 
which has interesting applications\cite{Cotsakis:1997cj,Querella:1998ke}. 

Furthermore, if one wants to write physical conservation laws in terms
of the covariant derivatives associated with the connection $\hat{\Gamma}$, 
the resulting equations seem to indicate violations of the matter energy
continuity and the equivalence principle\cite{Hamity:1992aa,Querella:1998ke}.
However, this is devoid of physical meaning - but the caveat is that one might also 
in the Palatini formalism switch to the Einstein frame where the metric 
$h_{\mu\nu}$ is physical. The two frames are not symmetric, which is manifest 
in the fact that the metric $g_{\mu\nu}$ continues to have some relevance 
in the Einstein frame: although the metric we measure there is $h_{\mu\nu}$, also there the 
continuity and the geodesics of matter are given by $g_{\mu\nu}$. In this 
sense we might say that Einstein frame theory furnishes a bi-metric 
structure, since there the measured metric is $h_{\mu\nu}$, but 
some aspects of the motion of matter are associated with the other metric 
$g_{\mu\nu}$. Consequently, the conservation of matter energy-momentum 
tensor as well as the geodesic hypothesis is violated in this frame. Note 
that all this happens also in the Einstein frame of extended gravities
in the metric formulation, and that such bi-metricity arises in fact in 
any other conformally equivalent frame except the Jordan one. Thus the
Palatini variational principle does not lead to new features when  
geodesics are concerned (assuming that the independent connection does 
not enter into the matter action).

This has not been very clear in the literature. It is often stated
that the Palatini formulation of extended gravity theories are bi-metric
in the sense that their metric structure\footnote{Sometimes this is also 
stated about the chronological structure. Our interpretation here is in 
any case simply that the metric structure (and the chronological 
structure, if it means the same thing) is given by the unique metric 
that is the physical one in the usual sense discussed above and for 
example in the Ref.\cite{Magnano:1993bd}. The 
causal structure is always left invariant by a conformal transformation, 
because it maps light-cones into light-cones (however, when the conformal 
transformation is not well defined, one might be lead into interesting 
situations by using so called conformal continuations.) Therefore there 
would not be a reason why the causal structure (or the chronological 
structure, if this would mean the same thing) would be preferably 
associated to a certain metric among the class of conformally equivalent 
metrics.} is determined by $g_{\mu\nu}$, while their geodesic structure 
would be given by the conformally equivalent metric $h_{\mu\nu}$ 
\cite{Allemandi:2004yx,Allemandi:2005tg,Poplawski:2005sc,Sotiriou:2005xe}. 
To highlight the difference of this notion to ours: 1) In the Einstein frame, 
we find exactly the opposite roles for the two metrics, 2) In the Jordan 
frame, the same metric $g_{\mu\nu}$ assumes both of the roles. It is 
always possible to introduce various connections 
on a manifold, but the one\footnote{Note that we are discussing actions 
(\ref{action}) and (\ref{action_p}). Our conclusions would not hold when
for example the matter Lagrangian would depend on some new metric.} we find
physically meaningful is selected by the motion of particles. The free fall 
of a particle follows the geodesics of this connection, which is also the 
Levi-Civita connection of the metric that minimizes the proper time 
integral of the particle along its path. Then the geodesic structure of 
the theory is due to this metric, at least to us an interpretation of any 
other definition of the geodesic structure would seem a bit vague.  

The metric and geodesic structures cannot be arbitrarily decoupled,  
since the free fall of particles is uniquely determined by the 
equations of motion for matter. These in turn, as shown in detail above, 
follow from the field equations which can be written solely in terms of 
the metric. Thus the laws governing the motion of particles are 
inscribed to the field equations, although they are occasionally 
introduced as a seemingly independent postulate. As a concrete example, 
consider dust. Then 
we can write $T_{\mu\nu} = \rho u_\mu u_\nu$, where $\rho$ is the energy 
density and $u_{\mu}$ the $n$-velocity of the fluid. The covariant 
conservation then gives
\be \label{dust}
\nabla^\mu T_{\mu\nu} = u_{\nu}\nabla^\mu(\rho u_\mu) + \rho u_\mu\nabla^\mu u_\nu = 0.
\ee
Because $u_\mu u^\mu = -1$, we find by multiplying this equation with 
$u_\nu$ that $\nabla^\mu(\rho u_\mu) = 0$. Therefore Eq.(\ref{dust}) 
reduces to $u_\mu \nabla^\mu u_\nu = 0$, which is nothing but the 
statement that the dust particles follow the geodesics of the metric 
$g_{\mu\nu}$. This already shows that geodesics in general cannot be 
determined by the $h_{\mu\nu}$-compatible connection $\hat{\Gamma}$ that 
appears in the action \ref{action_p}. The generalization of the above 
derivation of the geodesic equation from the covariant conservation of 
matter energy-momentum to the case of an arbitrary body of sufficiently 
small size and mass turns out to be much less trivial, and for this more 
general case we refer the reader to \cite{Ehlers:2003tv}. 
 
\subsection{Conclusion and outlook}
\label{i_c}

We have shown that the covariant conservation of energy-momentum and the
geodesic hypothesis continue to hold in extended gravity theories,
both in their metric and the Palatini form. This explicitly verified the  
self-consistency and geodesic uniqueness of these theories, thus 
possibly providing some clarifications, especially about the Palatini 
variation currently discussed in the literature. We also reviewed 
some essential earlier criticisms of the first order formalism.

With the understanding that when subjected to the ''Palatini's device'',
the $f(R)$ gravities exhibit a rather minimal modification of general 
relativity, they might seem a more appealing alternative to dark 
matter and dark energy. As the geodesic structure and the metric 
structure are entangled in the same way as in Einstein gravity and the 
equations of motion are still of 
the second order, we are not lead to fundamentally different concepts, we 
just have a modified coupling of matter to gravity. Since the 
gravitational field equations have not been experimentally tested at 
so vast scales, they might indeed deviate from the Einstein theory at 
cosmological distances. 

In practise, construction of a viable gravitational alternative to dark 
energy by extending the gravitational action has proven to be 
intricate. The problems always have to do with derivatives.
For higher-derivative gravity in the metric formulation, there is the Cauchy
issue mentioned in section \ref{m_b}, and also the doubt whether
the Ostrogradski theorem\cite{Woodard:2006nt} can admit
anything more general than $f(R)$ models. Furthermore, these models have been 
shown to feature ghosts and instabilities. All these are consequences of 
the higher derivatives present in these theories when compared to general 
relativity. Therefore these problems are avoided in the Palatini 
formulation. However, in that case there appears an extra double 
derivation, not effectively in the gravitational sector,
but in the matter sector. Matter is coupled to gravity via additional 
(covariant) derivatives of the trace of $T_{\mu\nu}$. When the 
idealization of a perfectly smooth universe is studied, this poses no 
difficulties, but by considering cosmological 
structures\cite{Koivisto:2005yc} one finds that the models 
in their present form are essentially ruled out\cite{Koivisto:2006ie} by
the observations of large scale structure when combined with the cosmic 
microwave background anisotropy measurements. 

A promising approach to build a consistent model of modified gravity which 
could viably account for the observed acceleration of the universe would 
be to find out ways to generalize Einstein equations without 
introducing higher derivatives either into the gravitational or into the 
matter section. 

\acknowledgments{The author is grateful to H. Kurki-Suonio for many useful 
suggestions about the manuscript. Section \ref{i_g} benefitted also from 
discussions with T. Sotiriou and G. Allemandi. This work was supported in 
part by NorFA, Waldemar von Frenckels Stiftelse and Emil Aaltosen 
S\"a\"ati\"o. During the preparation of the second version of the 
manuscript the author was funded by the Magnus Ehrnrooth Foundation.}

\bibliography{refs}

\end{document}